\documentclass[aps,prd,superscriptaddress,twocolumn]{revtex4-1}

\usepackage{amsfonts}
\usepackage{amsmath}
\usepackage{graphicx}
\usepackage{xcolor}
\usepackage{verbatim}
\usepackage[dvips]{epsfig}
\usepackage{float}
\usepackage{epsfig,subfig}
\usepackage{epstopdf}
\usepackage{hyperref}
\usepackage{cleveref}
\usepackage{placeins}

\begin{document}

\title{Quantum key-distribution protocols based on a quantum\\ version of the Monty Hall game}

\author{L. F. Quezada}
\email{luis.fernando@correo.nucleares.unam.mx}
\affiliation{Centro de Innovaci\'on y Desarrollo Tecnol\'ogico en C\'omputo, Instituto Polit\'ecnico Nacional, UPALM, 07700, Ciudad de M\'exico, M\'exico.}

\author{Shi-Hai Dong}
\email{dongsh2@yahoo.com}
\affiliation{Centro de Innovaci\'on y Desarrollo Tecnol\'ogico en C\'omputo, Instituto Polit\'ecnico Nacional, UPALM, 07700, Ciudad de M\'exico, M\'exico.}

\begin{abstract}
This work shows a possible application of quantum game theory to the area of quantum information, in particular to quantum cryptography. Here, we proposed two quantum key-distribution (QKD) protocols based on the quantum version of the Monty Hall game devised by Flitney and Abbott in \cite{QMH2}. Unlike most QKD protocols, in which the bits from which the key is going to be extracted are encoded in a basis choice (as in BB84), we encode these in an operation choice. The first proposed protocol uses qutrits to describe the state of the system and the same game-operators as in \cite{QMH2}. The motivation behind the second proposal was to simplify a possible physical implementation by adapting the formalism of the qutrit protocol to use qubits and simple logical quantum gates. In both protocols the security relies on the violation of a Bell-type inequality, for two qutrits and for six qubits in each case. Results show a higher ratio of violation than the E91 protocol.
\end{abstract}

\maketitle

\section{Introduction}

In 1926, Vernam invented the one-time pad encryption \cite{OneTimePad}, which uses a random secret key shared between two parties to encrypt a message. More than two decades later, in 1949, Shannon proved that the one-time pad scheme is optimal \cite{Shannon}, provided that the key is not reused by the parties. Therefore, in order to implement this scheme, the communicating parties must have a secure method to generate and share a random key that is as long as the message to be encrypted.

Consequently, one of the main goals of quantum cryptography nowadays, is to build a methodology that allows two parties to share a secure random key by taking advantage of the properties of quantum systems. To date, a variety of quantum key-distribution (QKD) protocols have been proposed for this task \cite{BB84, BB92, 6S_1, 6S_2, SARG04, E91, DPS, KMB09}.

Of particular relevance are the protocols developed by Charles Bennett and Gilles Brassard in 1984 (BB84) \cite{BB84}, and by Artur Ekert in 1991 (E91) \cite{E91}, as they were the first proposals and each one of them uses a different property of quantum systems to securely accomplish the key distribution. While the security of the BB84 protocol relies on the Heisenberg's uncertainty principle, the security of the E91 protocol is grounded in the non-classical correlations that arise between quantum entangled systems, which are usually tested using Bell-type inequalities \cite{Bell0,Bell1,Bell2,Bell3,Bell4,Bell5,Bell6}.

On the other hand, the area of mathematics known as game theory, found one of its main applications in secure classical communications \cite{GTCC1,GTCC2,GTCC3}, as eavesdropping can be treated as a game in which the spy's goal is to extract the maximum amount of information from a communication channel. This motivated quantum information theorists to begin including elements of quantum theory such as superposition of classical states and quantum entanglement into classical game theory \cite{QG1,QG2,QG3,QG4,QG5,QG6}, creating what is now known as quantum game theory.

One of the games that caught the attention of quantum theorists was the so-called Monty Hall game, which with its counter-intuitive result, led to a great debate between some mathematicians and probability experts in the 1970s and 1990s \cite{CMH1,CMH2,CMH3}. To date, due to the fact that the quantization procedure of a classical game is an entirely subjective task, there are various quantization schemes of the Monty Hall game \cite{QMH1,QMH2,QMH3,QMH4,QMH5,QMH6}. The most relevant scheme for the purposes of this paper is the one developed by Flitney and Abbott \cite{QMH2}.

In this work we develop a QKD protocol using qutrits, and based on Flitney and Abbott's quantization scheme of the Monty Hall game. Unlike most QKD protocols, in which the bits from which the key is going to be extracted are encoded in a basis choice (as in BB84), we encode these in an operation choice. The security of our protocol relies on the violation of a Bell-type inequality for two qutrits. Furthermore, in order to simplify a possible physical implementation, we also construct an analogue protocol using qubits and simple quantum gates, making it feasible to run on nowadays' quantum machines. In this case the security of the protocol relies on the violation of a Bell-type inequality for six qubits.

The paper is organized as follows. After a brief summary of the classical Monty Hall game in Sec. \ref{cMH}, we give an outline of the quantization scheme of the Monty Hall game devised by Flitney and Abbott in Sec. \ref{Fli_Abb}. Sec. \ref{qutrits} corresponds to the proposed protocol using qutrits and has three subsections: Subsec. \ref{QHKD}, where the protocol is described, Subsec. \ref{SecBt}, where the security of the protocol is grounded, and Subsec. \ref{Evet}, where the possible vulnerabilities of the protocol are addressed. Sec. \ref{qubits} corresponds to the proposed protocol using qubits and has three analogous subsections: Subsec. \ref{QHKDb}, Subsec. \ref{SecBb} and Subsec. \ref{Eveb}.

\section{Brief summary of the classical Monty Hall game} \label{cMH}

The Monty Hall game is a famous, seemingly paradoxical problem in probability \cite{CMH1,CMH2, CMH3}. It describes a contest in which a player is asked to choose between three doors, behind one of which a prize was randomly placed beforehand. There are two main characters in this contest: the host (Monty Hall), who knows behind which door the prize is, and the player, who does not have any information about its location.

The contest begins with the player choosing (but not opening) one of the doors. If the chosen door is the one with the prize behind, the host, who knows where the prize hides, randomly opens one of the two empty doors. On the other hand, if the player chooses one of the empty doors, the host opens the other remaining empty door. In both cases the host shares this information with the player. Lastly, the host asks the player if he wants to open his initial choice or prefers to open the other door that remains closed. The apparent paradox results from the fact that, when doing the calculations, it is found that the probability of the player finding the prize behind the door he initially chose is $1/3$, while the probability of finding the prize if he decides to open the other door is $2/3$.

\section{Flitney and Abbott's quantization scheme of the Monty Hall game \cite{QMH2}} \label{Fli_Abb}

In their article \cite{QMH2}, Flitney and Abbott use the classical characters of quantum information: ``Alice'' as the Host and ``Bob'' as the player. A state of the Monty Hall game is then represented as
\begin{equation}
\left| \psi \right\rangle = \left| o \, b\, a \right\rangle,
\end{equation}
where $a \in \left\lbrace 0,1,2 \right\rbrace $ is the door behind which Alice initially hid the prize, $b \in \left\lbrace 0,1,2 \right\rbrace $ is the door chosen (not opened) by Bob and $o \in \left\lbrace 0,1,2 \right\rbrace $ is the empty door to be opened. The initial state of the game is labeled as $\left| \psi_{i} \right\rangle$ and the final state $\left| \psi_{f} \right\rangle$ is calculated as
\begin{equation}
\left| \psi_{f} \right\rangle = \left( \cos\gamma \, \hat{S} + \sin\gamma \, \hat{I}_{27} \right) \, \hat{O} \, \left( \hat{I}_{3} \otimes \hat{B} \otimes \hat{A} \right)  \left| \psi_{i} \right\rangle,
\end{equation}
where $\hat{A}$ is Alice's strategy and determines the state of the hidden prize. $\hat{B}$ is Bob's strategy or choice-of-door operator. $\hat{O}$ is the empty-door-opening operator and selects a door to be opened depending on the value of $a$ and $b$. $\hat{S}$ is the door-switching operator and selects a different door depending on the values of $b$ and $o$. $\hat{I}_{n}$ is the identity operator of dimension $n$ and $\gamma \in \left\lbrace 0, \frac{\pi}{2} \right\rbrace $ depending on Bob's willingness to apply the switching operator. It is worth mentioning that $\hat{A}$, $\hat{B}$, $\hat{O}$ and $\hat{S}$ are all special unitary operators.

The empty-door-opening operator is defined as
\begin{equation}
\hat{O} = \displaystyle\sum_{i j k \ell} \left| \epsilon_{i j k}  \right| \,   \left| n j k \right\rangle \left\langle \ell j k \right| + \displaystyle\sum_{j \ell} \left| m j j \right\rangle \left\langle \ell j j \right|, \label{O}
\end{equation}
where $\epsilon_{i j k}$ is the Levi-Civita symbol, $m = (j+\ell+1)$ (mod 3) and $n = (i+\ell)$ (mod 3).

The door-switching operator is defined as
\begin{equation}
\hat{S} = \displaystyle\sum_{i j k \ell} \left| \epsilon_{i j \ell}  \right| \,   \left| i \ell k \right\rangle \left\langle i j k \right| + \displaystyle\sum_{i j} \left| i i j \right\rangle \left\langle i i j \right|, \label{S}
\end{equation}
where, as mentioned by Flitney and Abbott, the second term is just added to ensure the unitarity of the operator, as it maps states in which the opened door is the same as Bob's chosen one, i.e. $o = b$, something that does not happen in the game ($\hat{O}$ is explicitly constructed  to avoid that case). Notice that both $\hat{O}$ and $\hat{S}$ map every basis state $\left| i \, j \, k \right\rangle$ to a unique basis state.

Under this quantization scheme, the probability of Bob winning the prize is given by 
\begin{equation}
\left\langle \$_{B} \right\rangle  = \displaystyle\sum_{i j} \left| \left\langle i j j | \psi_{f} \right\rangle \right|^{2},
\end{equation}
and one can recover the classical result by choosing, for example
\begin{equation}
\left| \psi_{i} \right\rangle = \left| 0 \, 0 \, 0 \right\rangle,
\end{equation}
\begin{equation}
\hat{A} = \hat{B} = \left(
\begin{array}{ccc}
\displaystyle\frac{1}{\sqrt{3}} \quad & 0 \quad & \sqrt{\displaystyle\frac{2}{3}} \\ \\
\displaystyle\frac{1}{\sqrt{3}} \quad & \displaystyle\frac{-1}{\sqrt{2}} \quad & \displaystyle\frac{-1}{\sqrt{6}} \\ \\
\displaystyle\frac{1}{\sqrt{3}} \quad & \displaystyle\frac{1}{\sqrt{2}} \quad & \displaystyle\frac{-1}{\sqrt{6}} \\
\end{array}
\right).
\end{equation}

\section{QKD protocol using qutrits} \label{qutrits}

\subsection{Decription of the protocol}\label{QHKD}
In this subsection we present a step-by-step description of a proposed QKD protocol based on the quantum Monty Hall (QMH) game devised by Flitney and Abbott \cite{QMH2}. Just as in QMH, the protocol considers two parties: Alice and Bob.

\begin{enumerate}
	\setcounter{enumi}{0}
	\item \textit{Alice generates the state}
	\begin{equation}
	\left| \psi_{i} \right\rangle = \left| 0 \right\rangle \otimes \frac{1}{\sqrt{3}}\left( \left| 00 \right\rangle +\left| 11 \right\rangle + \left| 22\right\rangle  \right) \label{psii}
	\end{equation}
	\textit{as the initial state of the game.}	
\end{enumerate}

It is worth mentioning the importance of $\left| \psi_{i} \right\rangle$ as the initial state, particularly of the GHZ state \cite{GHZ} in which the first two qutrits are prepared. As noted by Benjamin and Hayden in \cite{QG3}, the GHZ state has a useful property in quantum game theory, that is
\begin{multline}
\left( \hat{U^{*}} \otimes \hat{U} \right) \frac{1}{\sqrt{3}}\left( \left| 00 \right\rangle +\left| 11 \right\rangle + \left| 22\right\rangle  \right) \\ = \frac{1}{\sqrt{3}}\left( \left| 00 \right\rangle +\left| 11 \right\rangle + \left| 22\right\rangle  \right),
\end{multline}
where, in this case, $\hat{U} \in$ SU($3$) and $\hat{U^{*}}$ stands for its complex conjugate. In the quantum game theory context, this property can be regarded as the existence of a counter-strategy $\hat{U^{*}}$ for the initially applied strategy $\hat{U}$.

Alice now needs to ``hide the prize''. This is represented in the following step.

\begin{enumerate}
	\setcounter{enumi}{1}
	\item \textit{Alice generates a random $n$-tuple of bits $k_{a}$ ($n$ will be the length of the raw key) and applies one of the following two operators depending on the bit in turn:}
	\begin{equation}
	\hat{G}_{0} = \left(
	\begin{array}{ccc}
	0 \quad & 0 \quad & 1 \\ \\
	1 \quad & 0 \quad & 0 \\ \\
	0 \quad & 1 \quad & 0 \\
	\end{array}
	\right) \quad \textrm{if the bit is $0$,}
	\end{equation}
	\begin{equation}
	\hat{G}_{1} = \left(
	\begin{array}{ccc}
	0 \quad & 1 \quad & 0 \\ \\
	0 \quad & 0 \quad & 1 \\ \\
	1 \quad & 0 \quad & 0 \\
	\end{array}
	\right) \quad \textrm{if the bit is $1$.}
	\end{equation}
\end{enumerate}

Notice that $\hat{G}_{0}$ and $\hat{G}_{1}$ act on the basis states $\left\lbrace \left| 0 \right\rangle, \left| 1 \right\rangle, \left| 2 \right\rangle  \right\rbrace $ as the sum and subtraction of $1$ (mod $3$) respectively.

\begin{enumerate}
	\setcounter{enumi}{2}
	\item \textit{Alice sends the second qutrit to Bob through a quantum channel.}
\end{enumerate}

Step 3 is one with a possible vulnerability, as Eve may be spying on the channel. We will address this and other security details of the protocol in the next subsections.

Bob now has to ``choose a door''. This is represented in the next step.

\begin{enumerate}
	\setcounter{enumi}{3}
	\item \textit{Bob generates a random $n$-tuple of bits $k_{b}$ ($n$ will be the length of the raw key) and applies $\hat{G}_{0}$ or $\hat{G}_{1}$ depending on the bit in turn.}
\end{enumerate}

The following step is also one with a possible vulnerability due to the presence of Eve.

\begin{enumerate}
	\setcounter{enumi}{4}
	\item \textit{Bob sends back his qutrit to Alice through a quantum channel.}
\end{enumerate}

To this phase of the protocol, both Alice and Bob have applied their strategies, and now Alice has the entire system in possession, meaning that every following operations will be performed by her.

Following QMH, Alice now has to ``open an empty door'', which is attained by applying the empty-door-opening operator \eqref{O}.

\begin{enumerate}
	\setcounter{enumi}{5}
	\item \textit{Alice applies $\hat{O}$ to the full state of the system.}
\end{enumerate}

Bob now has to ``choose'' if he wishes to switch between doors or stay with his initial choice. In the QMH context, this corresponds to ``choose'' between applying the door-switching operator \eqref{S} ($\gamma = 0$) and applying an identity operator ($\gamma = \frac{\pi}{2}$).

\begin{enumerate}
	\setcounter{enumi}{6}
	\item \textit{Bob generates a random $n$-tuple of bits $k_{s}$ that encodes with $0$ the case in which Alice will apply the door-switching operator \eqref{S} and with $1$ the case in which she won't. Bob makes $k_{s}$ public.}
\end{enumerate}

Step 7 may also be one with a possible vulnerability, as Eve is supposed to know every public information.

\begin{enumerate}
	\setcounter{enumi}{7}
	\item \textit{Depending on the in-turn bit of $k_{s}$, Alice applies the door-switching operator \eqref{S} (bit 0) or does nothing (bit 1).}
\end{enumerate}

For the purpose of this protocol, another operator is needed. We define the \textit{victory-encoding} operator:
\begin{equation}
\hat{V} = \displaystyle\sum_{i j k} \left| m j k \right\rangle \left\langle i j k \right|, \label{V}
\end{equation}
where $m = (i+j+k)$ (mod 3). As its name suggests, $\hat{V}$ encodes if Bob has win or not in the third qutrit. It is worth mentioning that $\hat{V}$ only acts as a victory-encoding operator for the states considered in this protocol, and not for an arbitrary state $\left| o \, b \, a \right\rangle$.

\begin{enumerate}
	\setcounter{enumi}{8}
	\item \textit{Alice applies the victory-encoding operator \eqref{V}.}
\end{enumerate}

To this stage of the protocol, the possible states in which the game can be, are given by
\begin{equation}
\left| \psi_{yx} \right\rangle  = \hat{V} \left( \cos\gamma \, \hat{S} + \sin\gamma \, \hat{I}_{27} \right) \, \hat{O} \, \left( \hat{I}_{3} \otimes \hat{G}_{y} \otimes \hat{G}_{x} \right)  \left| \psi_{i} \right\rangle,
\end{equation}
where $x,y \in \left\lbrace 0,1 \right\rbrace$ represent the choice of $\hat{G}_{0}$ or $\hat{G}_{1}$ by Alice and Bob, and $\gamma \in \left\lbrace 0,\frac{\pi}{2} \right\rbrace$ depending on the in-turn bit of $k_{s}$. For clarity in the final steps of the protocol, we breakdown these states:
\begin{align}
\left| \psi_{00} \right\rangle = \left| \psi_{11} \right\rangle &= \cos\gamma \left[ \displaystyle\frac{1}{\sqrt{3}} \left( \left| 001 \right\rangle + \left| 012 \right\rangle + \left| 020 \right\rangle  \right) \right] \notag
\\ & + \sin\gamma \left[ \displaystyle\frac{1}{\sqrt{3}} \left( \left| 100 \right\rangle + \left| 111 \right\rangle + \left| 122 \right\rangle  \right)  \right], \label{psi00}
\end{align}
\begin{align}
\left| \psi_{01} \right\rangle &= \cos\gamma \left[ \displaystyle\frac{1}{\sqrt{3}} \left( \left| 100 \right\rangle + \left| 111 \right\rangle + \left| 122 \right\rangle  \right) \right] \notag
\\ & + \sin\gamma \left[ \displaystyle\frac{1}{\sqrt{3}} \left( \left| 001 \right\rangle + \left| 012 \right\rangle + \left| 020 \right\rangle  \right)  \right], \label{psi01}
\end{align}
\begin{align}
\left| \psi_{10} \right\rangle &= \cos\gamma \left[ \displaystyle\frac{1}{\sqrt{3}} \left( \left| 200 \right\rangle + \left| 211 \right\rangle + \left| 222 \right\rangle  \right) \right] \notag
\\ & + \sin\gamma \left[ \displaystyle\frac{1}{\sqrt{3}} \left( \left| 002 \right\rangle + \left| 010 \right\rangle + \left| 021 \right\rangle  \right)  \right]. \label{psi10}
\end{align}

Notice from these expressions that in all the cases where the first two qutrits coincide ($b=a$), which corresponds to the cases in which Bob wins the game, the third qutrit is different from $\left| 0 \right\rangle$; while in the cases where the first two qutrits do not coincide ($b\neq a$), which corresponds to the cases in which Bob loses the game, the third qutrit is $\left| 0 \right\rangle$.

\begin{enumerate}
	\setcounter{enumi}{9}
	\item \textit{Alice measures the third qutrit. If the result is $\left| 0 \right\rangle$, it means $b \neq a$ and thus Bob has lost. If the result is either $\left| 1 \right\rangle$ or $\left| 2 \right\rangle$, it means $b = a$ and thus Bob has won.}
\end{enumerate}

Just as step 7, the next step may also represent a vulnerability due to the fact that Eve is supposed to know every public information.

\begin{enumerate}
	\setcounter{enumi}{10}
	\item \textit{Alice encodes the result of the game in a bit: $0$ if Bob lost and $1$ if Bob won. With many of these bits from various games, Alice forms the $n$-tuple $k_{r}$, which she then makes public.}
\end{enumerate}

From the states in \eqref{psi00}, \eqref{psi01} and \eqref{psi10}, notice that the cases in which Bob won by choosing not to switch between doors, were the ones where both Alice and Bob had chosen the same strategy $\hat{G}_{i}$. On the other hand, the cases in which Bob won by choosing to switch between doors, were the ones where Alice and Bob had chosen different strategies $\hat{G}_{i}$.

\begin{enumerate}
	\setcounter{enumi}{11}
	\item \textit{Alice and Bob publicly compare $k_{s}$ and $k_{r}$ from steps 7 and 11. If two bits in the same position (corresponding to the same game) coincide between $k_{s}$ and $k_{r}$, they also will coincide between $k_{a}$ and $k_{b}$. While if two bits do not coincide between $k_{s}$ and $k_{r}$, they also won't coincide between $k_{a}$ and $k_{b}$. In this last case Bob negates the corresponding bit of $k_{b}$. At the end of this process $k_{a} = k_{b}$.}
\end{enumerate}

This step allows Alice and Bob to use the $n$-tuples $k_{a}$ and $k_{b}$ as the key after performing an information-reconciliation process. There is one last step that has to do with the security of the protocol, this will be described along with its technical details in the next subsection.

\subsection{Security of the protocol}\label{SecBt}

In this subsection we describe the technical details in which the security of the protocol is based. To avoid introducing more notation, we will use $k_{i}$ to refer to either the tuple or just one of the bits of that tuple.

The tenth step of the protocol states that Alice must perform a measurement on the third qutrit in order for her to know if Bob has won or lost. With this measurement, the three-qutrit states in equations \eqref{psi00}, \eqref{psi01} and \eqref{psi10} will collapse into a two-qutrit subspace. After the measurement, the new state of the system depends on whether Alice and Bob applied the same strategy or not. However, as we will show next, just from the publicly available information ($k_{s}$ and $k_{r}$) and her own knowledge ($k_{a}$), Alice can know for sure which state remains after the measurement. 

\begin{itemize}
	\item In the cases where $\left\lbrace k_{a} = 0, k_{s} = 0, k_{r} = 1 \right\rbrace$ or $\left\lbrace k_{a} = 0, k_{s} = 1, k_{r} = 1 \right\rbrace$ or $\left\lbrace k_{a} = 1, k_{s} = 0, k_{r} = 1 \right\rbrace$ or $\left\lbrace k_{a} = 1, k_{s} = 1, k_{r} = 1 \right\rbrace$, the state of the system after the measurement is given by
\begin{equation}
\left| \phi_{0} \right\rangle = \displaystyle\frac{1}{\sqrt{3}} \left( \left| 00 \right\rangle + \left| 11 \right\rangle + \left| 22 \right\rangle  \right). \label{phi0}
\end{equation}

	\item In the cases where $\left\lbrace k_{a} = 0, k_{s} = 0, k_{r} = 0 \right\rbrace$ or $\left\lbrace k_{a} = 1, k_{s} = 1, k_{r} = 0 \right\rbrace$ or $\left\lbrace k_{a} = 1, k_{s} = 0, k_{r} = 0 \right\rbrace$, the state of the system after the measurement is given by
\begin{equation}
\left| \phi_{1} \right\rangle = \displaystyle\frac{1}{\sqrt{3}} \left( \left| 01 \right\rangle + \left| 12 \right\rangle + \left| 20 \right\rangle  \right). \label{phi1}
\end{equation}

	\item In the case where $\left\lbrace k_{a} = 0, k_{s} = 1, k_{r} = 0 \right\rbrace$, the state of the system after the measurement is given by
\begin{equation}
\left| \phi_{2} \right\rangle = \displaystyle\frac{1}{\sqrt{3}} \left( \left| 02 \right\rangle + \left| 10 \right\rangle + \left| 21 \right\rangle  \right). \label{phi2}
\end{equation}
\end{itemize}

Notice that $\left| \phi_{0} \right\rangle$, $\left| \phi_{1} \right\rangle$ and $\left| \phi_{2} \right\rangle$ are all entangled states. In fact $\left| \phi_{0} \right\rangle$ is the GHZ state for two three-dimensional parties, while both $\left| \phi_{1} \right\rangle$ and $\left| \phi_{2} \right\rangle$ are equivalent to $\left| \phi_{0} \right\rangle$ in the sense that both can be obtained from it via local operations. This is an important feature, as these states strongly (almost maximally) violate a Bell-type inequality for two qutrits \cite{Bell2,Bell3,Bell6}.

For two three-dimensional systems, every local hidden-variable (LHV) theory or classically correlated system must satisfy the following inequality \cite{Bell2,Bell3,Bell6}:
\begin{align}
I_{3} =& \, P\left( A_{1} = B_{1} \right) + P\left( A_{2} \oplus 1 = B_{1} \right) \notag
\\
&+ P\left( A_{2} = B_{2} \right) + P\left( A_{1} = B_{2} \right) \notag
\\
&- P\left( A_{1} \oplus 1 = B_{1} \right) - P\left( A_{2} = B_{1} \right) \notag
\\
&- P\left( A_{2} \oplus 1 = B_{2} \right) - P\left( A_{1} = B_{2} \oplus 1 \right) \leq 2,  \label{Bellt}
\end{align}
where $\oplus$ is the sum (mod 3) and $A_{i}, B_{i} \in \left\lbrace 0 ,1 ,2 \right\rbrace $ denote the three possible outcomes of two different measurements ($i \in \left\lbrace 1, 2\right\rbrace $) made in systems $A$ and $B$ respectively.

Notice that, quantum-mechanically, the value of $I_{3}$ in \eqref{Bellt} depends on the specific measurements the parties perform, as well as on the state in which the system is. As it is usual when working with Bell-type inequalities, we calculate the value of $I_{3}$ taking the expectation value of a Bell operator $\mathcal{B}$ with respect to the state of the system \cite{Bell3,Bell6}, i.e. $I_{3} = \left\langle \mathcal{B} \right\rangle $.

The Bell operators which maximally violate the inequality \eqref{Bellt} for the states $\left| \phi_{0} \right\rangle$, $\left| \phi_{1} \right\rangle$ and $\left| \phi_{2} \right\rangle$, are respectively:
\begin{equation}
\mathcal{B}_{0} = \left(
\begin{array}{ccccccccc}
0 & 0 & 0 & 0 & \frac{2}{\sqrt{3}} & 0 & 0 & 0 & 2 \\
0 & 0 & 0 & 0 & 0 & \frac{2}{\sqrt{3}} & 0 & 0 & 0 \\
0 & 0 & 0 & 0 & 0 & 0 & 0 & 0 & 0 \\
0 & 0 & 0 & 0 & 0 & 0 & 0 & \frac{2}{\sqrt{3}} & 0 \\
\frac{2}{\sqrt{3}} & 0 & 0 & 0 & 0 & 0 & 0 & 0 & \frac{2}{\sqrt{3}} \\
0 & \frac{2}{\sqrt{3}} & 0 & 0 & 0 & 0 & 0 & 0 & 0 \\
0 & 0 & 0 & 0 & 0 & 0 & 0 & 0 & 0 \\
0 & 0 & 0 & \frac{2}{\sqrt{3}} & 0 & 0 & 0 & 0 & 0 \\
2 & 0 & 0 & 0 & \frac{2}{\sqrt{3}} & 0 & 0 & 0 & 0 \\
\end{array}
\right),
\end{equation}
\begin{equation}
\mathcal{B}_{1} = \left(
\begin{array}{ccccccccc}
0 & 0 & 0 & 0 & 0 & 0 & 0 & 0 & 0 \\
0 & 0 & 0 & 0 & 0 & \frac{2}{\sqrt{3}} & 2 & 0 & 0 \\
0 & 0 & 0 & \frac{2}{\sqrt{3}} & 0 & 0 & 0 & 0 & 0 \\
0 & 0 & \frac{2}{\sqrt{3}} & 0 & 0 & 0 & 0 & 0 & 0 \\
0 & 0 & 0 & 0 & 0 & 0 & 0 & 0 & \frac{2}{\sqrt{3}} \\
0 & \frac{2}{\sqrt{3}} & 0 & 0 & 0 & 0 & \frac{2}{\sqrt{3}} & 0 & 0 \\
0 & 2 & 0 & 0 & 0 & \frac{2}{\sqrt{3}} & 0 & 0 & 0 \\
0 & 0 & 0 & 0 & 0 & 0 & 0 & 0 & 0 \\
0 & 0 & 0 & 0 & \frac{2}{\sqrt{3}} & 0 & 0 & 0 & 0 \\
\end{array}
\right),
\end{equation}
\begin{equation}
\mathcal{B}_{2} = \left(
\begin{array}{ccccccccc}
0 & 0 & 0 & 0 & \frac{2}{\sqrt{3}} & 0 & 0 & 0 & 0 \\
0 & 0 & 0 & 0 & 0 & 0 & 0 & 0 & 0 \\
0 & 0 & 0 & \frac{2}{\sqrt{3}} & 0 & 0 & 0 & 2 & 0 \\
0 & 0 & \frac{2}{\sqrt{3}} & 0 & 0 & 0 & 0 & \frac{2}{\sqrt{3}} & 0 \\
\frac{2}{\sqrt{3}} & 0 & 0 & 0 & 0 & 0 & 0 & 0 & 0 \\
0 & 0 & 0 & 0 & 0 & 0 & \frac{2}{\sqrt{3}} & 0 & 0 \\
0 & 0 & 0 & 0 & 0 & \frac{2}{\sqrt{3}} & 0 & 0 & 0 \\
0 & 0 & 2 & \frac{2}{\sqrt{3}} & 0 & 0 & 0 & 0 & 0 \\
0 & 0 & 0 & 0 & 0 & 0 & 0 & 0 & 0 \\
\end{array}
\right).
\end{equation}
The three of them yield a value of $I_{3} = \frac{4}{9} \left( 3 + 2 \sqrt{3} \right) \approx 2.873$.

It is worth mentioning that the ratio of violation for this inequality, defined in general as 
\begin{equation}
r = \displaystyle\frac{\left\langle \mathcal{B} \right\rangle_{QM}}{\left\langle \mathcal{B} \right\rangle_{LHV}},
\end{equation}
using the Bell operators $\mathcal{B}_{0}$, $\mathcal{B}_{1}$ and $\mathcal{B}_{2}$ with the states $\left| \phi_{0} \right\rangle$, $\left| \phi_{1} \right\rangle$ and $\left| \phi_{2} \right\rangle$ respectively, is $r = \frac{2.873}{2} \approx 1.435$; higher than the ratio of violation in the $E91$ protocol ($r=\sqrt{2} \approx 1.414$), which is the one given by the CHSH inequality violation \cite{Bell0}.

We are now in a position to describe the last step of the protocol.

\begin{enumerate}
	\setcounter{enumi}{12}
	\item \textit{Alice measures the states $\left| \phi_{j} \right\rangle$ and calculates the expectation value of the corresponding Bell operator $\mathcal{B}_j$ in each case. If $\left\langle \mathcal{B}_j \right\rangle \geq \chi$, where $\chi >  2$ is a previously agreed lower bound for $I_{3}$ between Alice and Bob, they conclude there was no interference in their communications and thus the key is safe.}
\end{enumerate}

\subsection{Eve's attack} \label{Evet}

In this subsection we describe how the presence of an spy (Eve) in the communications between Alice and Bob, might alter the results of the protocol proposed in subsection \ref{QHKD}. We suppose that every measurement performed by Eve is a projective measurement.

We assert that there is no vulnerability in the publicly available information by itself, namely $k_{s}$ and $k_{r}$; of course this is only the case when Eve just knows $k_{s}$ and $k_{r}$. This assertion can be easily proved by checking all possible combinations of values between $k_{s}$ and $k_{r}$, and noticing that in all of them, the applied strategies, which are the ones that encode the key, can not be unambiguously determined:

\begin{itemize}
	\item If $k_{s} = k_{r} = 0$, then the possible values for $k_{a}$ and $k_{b}$ are $\left\lbrace k_{a}=0, \, k_{b}=0  \right\rbrace$ or $\left\lbrace k_{a}=1, \, k_{b}=1  \right\rbrace$.
	
	\item If $k_{s} = 0, \, k_{r} = 1$, then the possible values for $k_{a}$ and $k_{b}$ are $\left\lbrace k_{a}=0, \, k_{b}=1  \right\rbrace$ or $\left\lbrace k_{a}=1, \, k_{b}=0  \right\rbrace$.
	
	\item If $k_{s} = 1, \, k_{r} = 0$, then the possible values for $k_{a}$ and $k_{b}$ are $\left\lbrace k_{a}=0, \, k_{b}=1  \right\rbrace$ or $\left\lbrace k_{a}=1, \, k_{b}=0  \right\rbrace$.
	
	\item If $k_{s} = k_{r} = 1$, then the possible values for $k_{a}$ and $k_{b}$ are $\left\lbrace k_{a}=0, \, k_{b}=0  \right\rbrace$ or $\left\lbrace k_{a}=1, \, k_{b}=1  \right\rbrace$.
\end{itemize}
However, as it can be seen from these cases, if Eve happens to know which strategy was applied by either Alice or Bob without them noticing, the whole protocol falls down.

The first and third qutrits are always in Alice's possession, meaning that Eve can not infer anything about Alice's applied strategy. However, the second qutrit, in which Bob applies his strategy, is sent through a quantum channel twice, meaning that Eve has two chances to hack the protocol.

The first possible case is that Eve intercepts and measures Bob's qutrit the first time it is sent (step 3 of the protocol), but not the second one. In this case, as Bob has not yet applied his strategy, there is no useful information Eve could retrieve from her measurement, getting only a qutrit in the state $\left| 0 \right\rangle$, $\left| 1 \right\rangle$ or $\left| 2 \right\rangle$ with a probability of $1/3$ each.

The second possible case is that Eve intercepts and measures Bob's qutrit the second time it is sent (step 5 of the protocol), but not the first one. This case might seem different from the last one, as this time Bob has already applied his strategy. However, due to the form of $\left| \psi_{i} \right\rangle$ in \eqref{psii}, Eve would again only be getting a qutrit in the state $\left| 0 \right\rangle$, $\left| 1 \right\rangle$ or $\left| 2 \right\rangle$ with a probability of $1/3$ each.

It is clear that in order for Eve to have a chance of knowing which strategy Bob will apply, she has to intercept the second qutrit both the first and the second time it is sent. The fist one to project the qutrit in an specific and arbitrary state known by her, and the second one to retrieve the information regarding the strategy applied by Bob; this kind of attack is known as an \textit{Intercept and Resend} or $IR$ attack . The restriction of Eve having to perform two measurements in order to gain information from the system may be exploited to increase the security of the protocol, by using two different one-way quantum channels for example, a feature that could also protect it from more general attacks.

We have shown that in this paradigm of projective measurements, the only possible option for Eve is to perform two different $IR$ attacks, possibly in two different quantum channels. We next describe in detail what happens if she does.

As mentioned in the previous subsection \ref{SecBt}, the only possible states that can remain after Alice's measurement of the third qutrit are $\left| \phi_{0} \right\rangle$, $\left| \phi_{1} \right\rangle$ and $\left| \phi_{2} \right\rangle$ \eqref{phi0} \eqref{phi1} \eqref{phi2}. In each case, if Eve decided to apply a $2$-$IR$ attack every time Alice and Bob implemented the protocol, the possible states that remain after Alice's last measurement are respectively
\begin{align}
\hat{\varrho}_{0} =& \frac{1}{3} \left( \left| 00 \right\rangle \left\langle 00 \right| + \left| 11 \right\rangle \left\langle 11 \right| + \left| 22 \right\rangle \left\langle 22 \right| \right),
\\
\hat{\varrho}_{1} =& \frac{1}{3} \left( \left| 01 \right\rangle \left\langle 01 \right| + \left| 12 \right\rangle \left\langle 12 \right| + \left| 20 \right\rangle \left\langle 20 \right| \right),
\\
\hat{\varrho}_{2} =& \frac{1}{3} \left( \left| 02 \right\rangle \left\langle 02 \right| + \left| 10 \right\rangle \left\langle 10 \right| + \left| 21 \right\rangle \left\langle 21 \right| \right),
\end{align}
which can be shown to yield a value of $I_{3} = \textrm{Tr}\left\lbrace \hat{\varrho}_{j} \mathcal{B}_{j}  \right\rbrace = 0 $ for all $j\in \left\lbrace 0,1,2 \right\rbrace $. This result, based on what was exposed in the previous subsection \ref{SecBt}, allows Alice and Bob to infer that Eve has interfered in their communications.

\begin{figure}[!t]
	\begin{centering}
		\includegraphics[width=0.9\linewidth]{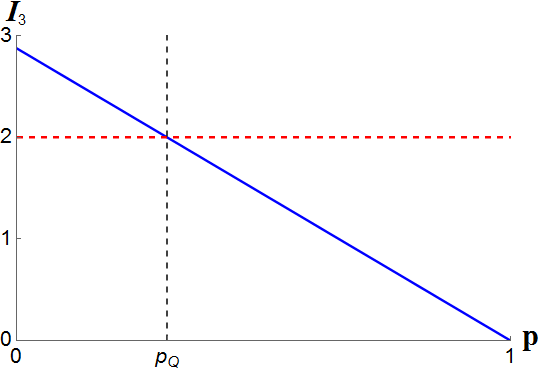}
		\par\end{centering}
	\caption{Dependence of the Bell parameter $I_{3}$ with the probability $p$ of Eve performing two $IR$ attacks. \label{figI3}}
\end{figure}

Eve knows that if she attacks the channel every time Alice and Bob implement the protocol, she is going to be detected. So she decides to carry out the $2$-$IR$ attack based on the occurrence of an event with probability $p$, i.e. Eve's attack will be executed with probability $p$ each time the protocol is implemented. Under these circumstances, the possible states that remain in each case after Alice's last measurement are
\begin{align}
\hat{\rho}_{0} =& \left( 1-p \right) \left| \phi_{0} \right\rangle \left\langle \phi_{0} \right| + p \, \hat{\varrho}_{0}, \label{rho0t}
\\
\hat{\rho}_{1} =& \left( 1-p \right) \left| \phi_{1} \right\rangle \left\langle \phi_{1} \right| + p \, \hat{\varrho}_{1}, \label{rho1t}
\\
\hat{\rho}_{2} =& \left( 1-p \right) \left| \phi_{2} \right\rangle \left\langle \phi_{2} \right| + p \, \hat{\varrho}_{2}. \label{rho2t}
\end{align}
In this case $I_{3}$ is a function of $p$, the dependence is the same for all $\hat{\rho}_{j}$ and its plot is shown in Figure \ref{figI3}.

Defining $p_Q$ as the value of $p$ in which $I_{3} = 2$, from Figure \ref{figI3} we found that if $p<p_{Q}$, the inequality \eqref{Bellt} can still be violated. The value of $p_Q$ is $p_Q = 11/2 - 3 \sqrt{3} \approx 0.304$, meaning that Eve can have approximately $30\%$ of the raw key if Alice and Bob decided that their only criterion to consider the key as safe was the inequality violation. However, as stated in step 13 of the protocol, Alice and Bob could have previously agreed on a lower bound for $I_{3}$, reducing Eve's information on the key as much as they want.

It is worth mentioning the effect that the presence of noise in the quantum channel has in the value of $I_{3}$. As it is shown in \cite{Bell3}, the presence of noise in the channel can be modeled in a similar fashion as the presence of Eve \eqref{rho0t} \eqref{rho1t} \eqref{rho2t}, meaning that in this case, some amount of noise in the channel plays against Eve, as even if she chooses a relatively small value of $p$, the presence of noise would amplify it, making it easier for Alice and Bob to detect her.

\section{QKD protocol using qubits} \label{qubits}
In order to motivate and simplify a possible physical implementation of the QKD protocol described in the previous section \ref{QHKD}, which we will refer to as the Quanty-Hall protocol, in this section we adequate its formalism to use qubits instead of qutrits, and simple quantum gates instead of unitary three-dimensional operators.

\subsection{Decription of the protocol}\label{QHKDb}
In this subsection we present a step-by-step description of the Quanty-Hall protocol, using qubits and simple quantum gates for its implementation. The translation is made by associating one qutrit with two qubits as:
\begin{align}
\left| 0 \right\rangle & \longrightarrow \left| 00 \right\rangle, \notag
\\
\left| 1 \right\rangle & \longrightarrow \left| 01 \right\rangle, \notag
\\
\left| 2 \right\rangle & \longrightarrow \left| 10 \right\rangle, \label{ass}
\end{align}
and ignoring the two-qubits state $\left| 11 \right\rangle$. With this association in mind, seems natural to think that we need just six qubits to model the system used in the Quanty-Hall protocol. However, due to the nature of the door-switching operator $\hat{S}$ \eqref{S}, which based on the values of $b$ and $o$ selects a different value of $b$, we need to add two ancillary qubits that will serve as control qubits for its application. It is worth mentioning that, despite the fact that the empty-door-opening operator $\hat{O}$ \eqref{O} has this same behavior, it is not necessary to add ancillary qubits for it; this is due to the fact that the value of qutrit $o$ is initialized in $o = 0$, and it remains with that value until the application of $\hat{O}$.

Unlike the qutrit-Quanty-Hall protocol, in this case we will suppose that the initial state of the system is with all eight needed qubits in zero, i.e.
\begin{equation}
\left| \psi_{bi} \right\rangle = \underbrace{\left| 0 0\right\rangle}_{B_{ns}} \otimes \underbrace{\left| 0 0\right\rangle}_{O} \otimes \underbrace{\left| 0 0\right\rangle}_{B_{s}} \otimes \underbrace{\left| 0 0\right\rangle}_{A}, \label{psiib}
\end{equation}
where the first two qubits ($A$) correspond to Alice's choice of strategy, the third and fourth qubits ($B_{s}$) correspond to Bob's choice of strategy and are the ones on which the door-switching operator will act, the fifth and sixth qubits ($O$) are the ones on which the information regarding the empty-door opening will be stored, and the seventh and eighth qubits ($B_{ns}$) are the ancillary qubits needed to control the door-switching operator. Qubits ($B_{ns}$) correspond to the state of Bob's qutrit in the case where he does not apply the door-switching operator, while qubits ($B_{s}$) correspond to the state of Bob's qutrit in the case where he does apply the door-switching operator. Due to the fact that in this case we have four qubits corresponding to Bob, and in order to avoid sending through a quantum channel redundant information, in this qubit-Quanty-Hall protocol, Bob is going to be the one performing most of the operations, while Alice will just apply its own strategy.

The first step of the protocol describes the operations that Bob must perform on the initial state \eqref{psiib} in order to analogously initialize the system as in the qutrit-Quanty-Hall protocol \eqref{psii}.

\begin{figure}[!t]
	\begin{centering}
		\includegraphics[width=0.9\linewidth]{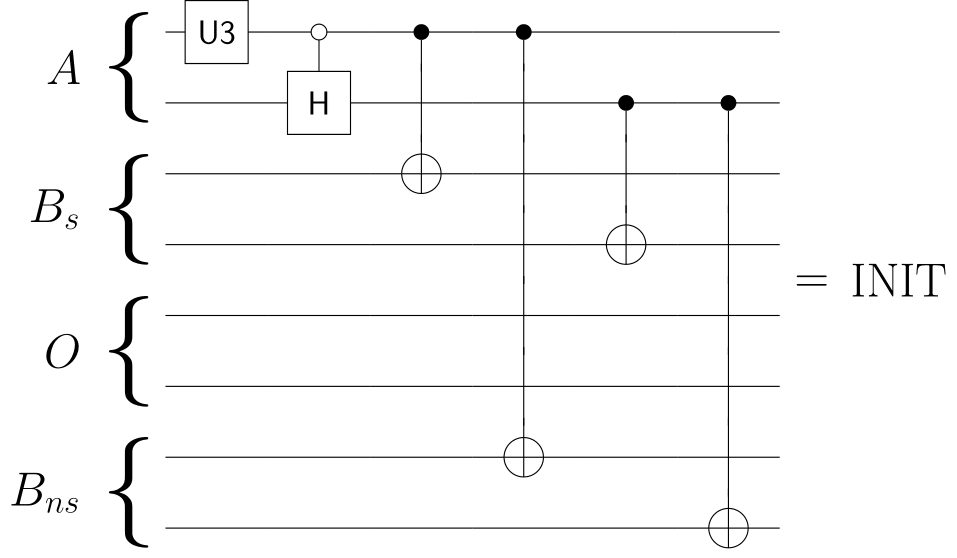}
		\par\end{centering}
	
	\caption{\label{fig:1} Quantum Circuit of the INIT operator. It initialize the system in the state $\frac{1}{\sqrt{3}}\left( \left| 00000000 \right\rangle +\left| 01000101 \right\rangle + \left| 10001010 \right\rangle \right)$.}
\end{figure}

\begin{enumerate}
	\setcounter{enumi}{0}
	\item \textit{Bob applies the INIT operator (see Figure \ref{fig:1}) to the initial state $\left| \psi_{bi} \right\rangle$.}
\end{enumerate}
Where $H$ in Figure \ref{fig:1} represents the Hadamard gate and $U3$ is the gate defined as
\begin{equation}
U3 = \left(
\begin{array}{cc}
\cos\left( \frac{\theta}{2} \right)  \quad & -e^{i\lambda} \sin\left( \frac{\theta}{2} \right) \\ \\
e^{i\phi} \sin\left( \frac{\theta}{2} \right) \quad & e^{i\lambda + i\phi} \cos\left( \frac{\theta}{2} \right)
\end{array}
\right)\Biggr\rvert \substack{\theta = 2 \arctan\left(\frac{1}{\sqrt{2}}\right)\\ \phi = 0 \\ \lambda = \pi.}
\end{equation}

The second step describes the application of Bob's strategy to his corresponding qubits.

\begin{figure}[!b]
	\begin{centering}
		\includegraphics[width=0.9\linewidth]{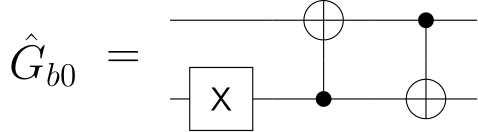}
		\par\end{centering}
		\caption{\label{fig:2} Quantum Circuit of the operator $\hat{G}_{b0}$.}
\end{figure}

\begin{figure}[!b]
	\begin{centering}
		\includegraphics[width=0.8\linewidth]{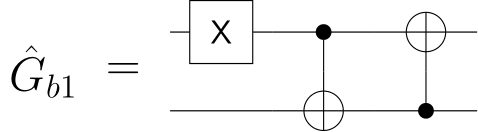}
		\par\end{centering}
		\caption{\label{fig:3} Quantum Circuit of the operator $\hat{G}_{b1}$.}
\end{figure}

\begin{enumerate}
	\setcounter{enumi}{1}
	\item \textit{Bob generates a random $n$-tuple of bits $k_{b}$ ($n$ will be the length of the raw key) and depending on the in-turn bit of $k_{b}$, applies $\hat{G}_{b0}$ or $\hat{G}_{b1}$ (see Figures \ref{fig:2} and \ref{fig:3}) to both pair of qubits, $B_{s}$ and $B_{ns}$, belonging to him.}
\end{enumerate}
where $X$ in Figures \ref{fig:2} and \ref{fig:3}  represent the Not gate.

\begin{figure}[!t]
	\begin{centering}
		\includegraphics[width=1.0\linewidth]{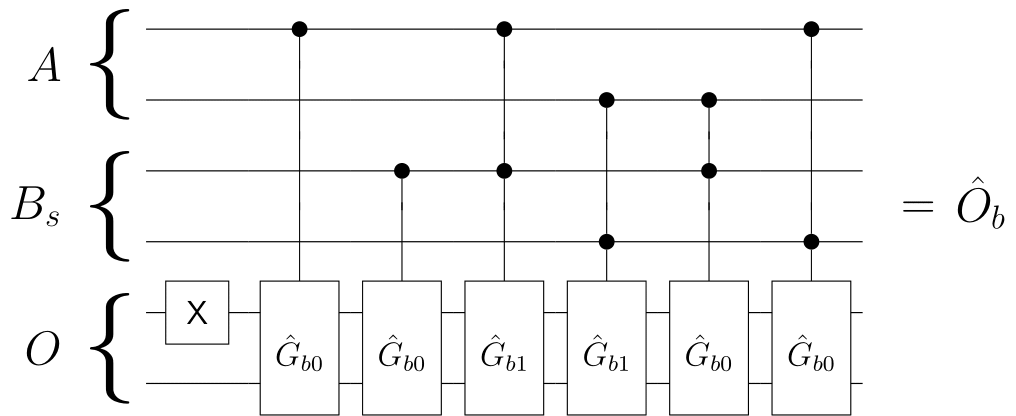}
		\par\end{centering}
		\caption{\label{fig:4} Quantum Circuit of the operator $\hat{O}_b$}
\end{figure}

For Alice to be able to apply her strategy, Bob must send her the corresponding pair of qubits.

\begin{enumerate}
	\setcounter{enumi}{2}
	\item \textit{Bob sends the pair of qubits $A$ to Alice through a quantum channel each.}
\end{enumerate}

With the pair of qubits $A$ in her possession, Alice now applies her strategy. 

\begin{enumerate}
	\setcounter{enumi}{3}
	\item \textit{Alice generates a random $n$-tuple of bits $k_{a}$ ($n$ will be the length of the raw key) and depending on the in-turn bit of $k_{a}$, applies $\hat{G}_{b0}$ or $\hat{G}_{b1}$ (see Figures \ref{fig:2} and \ref{fig:3}) to her pair of qubits.}
\end{enumerate}

In order for Bob to apply the remaining operations, Alice must return her pair of qubits to Bob.

\begin{enumerate}
	\setcounter{enumi}{4}
	\item \textit{Alice sends back the pair of qubits $A$ to Bob through a quantum channel each.}
\end{enumerate}

Now that Bob is in possession of all the state, he will be the one performing the remaining operations.

\begin{enumerate}
	\setcounter{enumi}{5}
	\item \textit{Bob applies $\hat{O}_{b}$ (see Figure \ref{fig:4}) to the first six qubits ($A$, $B_{s}$ and $O$).}
\end{enumerate}
Notice that, since Bob's qubits are only for controlling the operations performed on $O$, there is no need to include the pair of qubits $B_{ns}$, as to this point, they carry the same information as the pair of qubits $B_{s}$.

It is worth mentioning that the qubit operator $\hat{O}_{b}$ only acts as the qutrit operator $\hat{O}$ for the states that appear in the protocol, and not for an arbitrary state $\left| o \, b \, a \right\rangle$ translated to qubits by the association in \eqref{ass}.

Unlike the qutrit-Quanty-Hall protocol, this time is Bob the one that will apply the switching operator, and will do it to the corresponding qubits no matter what.

\begin{figure}[!b]
	\begin{centering}
		\includegraphics[width=1.0\linewidth]{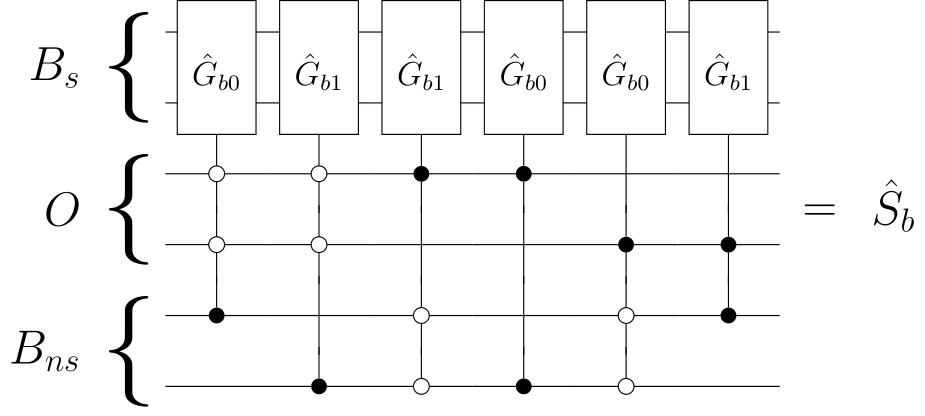}
		\par\end{centering}
	\caption{\label{fig:5} Quantum Circuit of the operator $\hat{S}_b$}
\end{figure}

\begin{enumerate}
	\setcounter{enumi}{6}
	\item \textit{Bob applies $\hat{S}_{b}$ (see Figure \ref{fig:5}) to the last six qubits ($B_{s}$, $O$ and $B_{ns}$).}
\end{enumerate}

As in the qutrit-Quanty-Hall protocol, we need again a victory-encoding operator.

\begin{figure}[!t]
	\begin{centering}
		\includegraphics[width=1.0\linewidth]{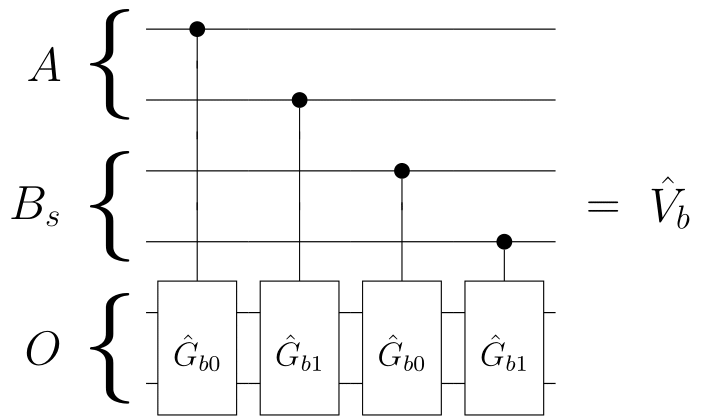}
		\par\end{centering}
		\caption{\label{fig:6} Quantum Circuit of the operator $\hat{V}_{b}$.}
\end{figure}

\begin{enumerate}
	\setcounter{enumi}{7}
	\item \textit{Bob applies $\hat{V}_{b}$ (see Figure \ref{fig:6}) to the first six qubits ($A$, $B_{s}$ and $O$).}
\end{enumerate}

We break down again all the possible states in which the system can be to this point of the qubit-Quanty-Hall protocol:
\begin{align}
\left| \psi_{b00} \right\rangle = \left| \psi_{b11} \right\rangle &= \overbrace{\left| 00 \right\rangle}^{O} \otimes \notag \\ &\displaystyle\frac{1}{\sqrt{3}}  \left( \left| 001000 \right\rangle + \left| 010001 \right\rangle + \left| 100110 \right\rangle  \right), \label{psib00}
\end{align}
\begin{align}
\left| \psi_{b01} \right\rangle &= \overbrace{\left| 01 \right\rangle}^{O} \otimes \notag \\ &\displaystyle\frac{1}{\sqrt{3}}  \left( \left| 000101 \right\rangle + \left| 011010 \right\rangle + \left| 100000 \right\rangle   \right), \label{psib01}
\end{align}
\begin{align}
\left| \psi_{b10} \right\rangle &= \overbrace{\left| 10 \right\rangle}^{O} \otimes \notag \\ &\displaystyle\frac{1}{\sqrt{3}}  \left( \left| 001010 \right\rangle + \left| 010000 \right\rangle + \left| 100101 \right\rangle   \right), \label{psib10}
\end{align}
where, for simplicity, we have rearranged the pairs of qubits from $\left| B_{ns}, \, O, \, B_{s}, \, A \right\rangle$ to $\left| O, \, B_{ns}, \, B_{s}, \, A \right\rangle$.

Notice from these expressions that the cases in which the results of the measurements of qubits $O$ are $00$, are the ones in which Alice and Bob had chosen the same strategy $\hat{G}_{bi}$; while the cases in which the results of the measurements of qubits $O$ are either $01$ or $10$, are the ones in which Alice and Bob had chosen different strategies $\hat{G}_{bi}$.

\begin{figure*}[t]
	\begin{centering}
		\includegraphics[width=1.0\linewidth]{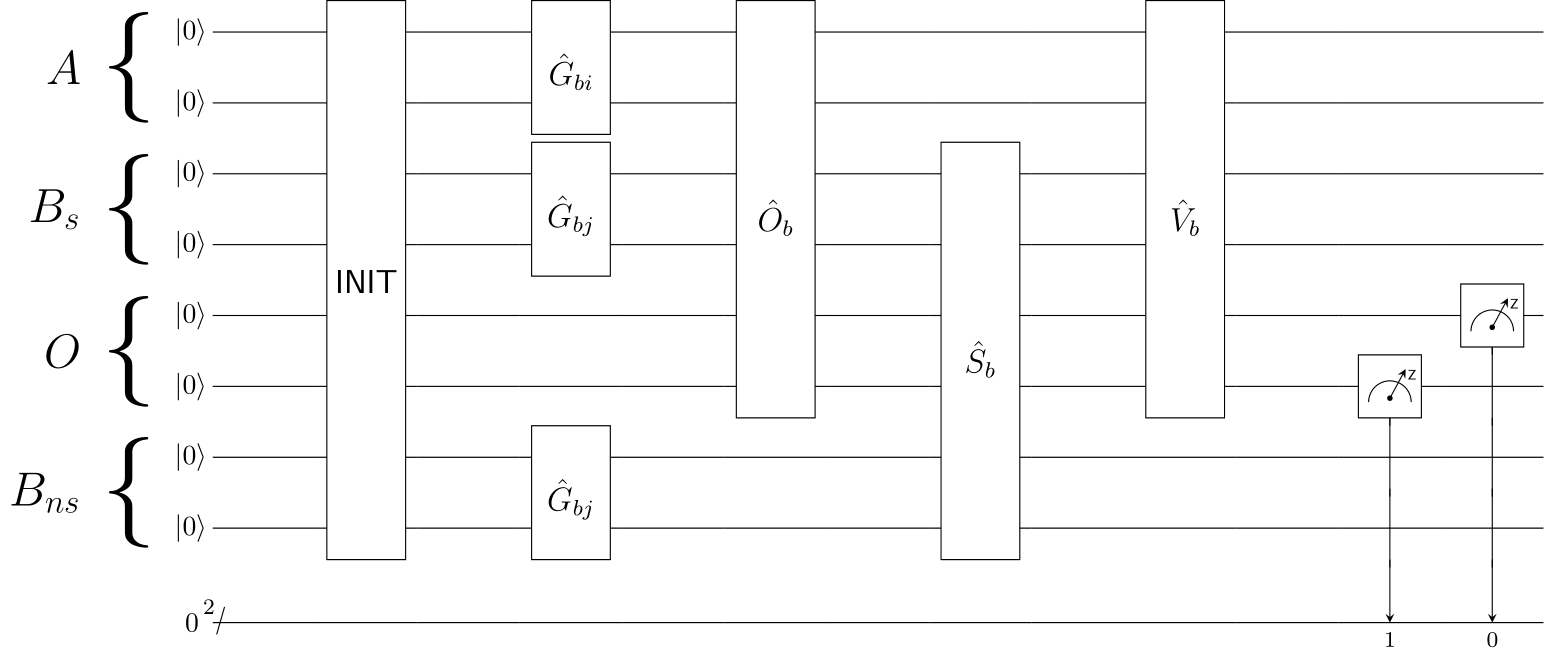}
		\par\end{centering}
	
	\caption{\label{fig:7} Quantum Circuit of the qubit-Quanty-Hall protocol.}
\end{figure*}

\begin{enumerate}
	\setcounter{enumi}{8}
	\item \textit{Bob measures the pair of qubits $O$. If the results are $00$, the corresponding bits between $k_{a}$ and $k_{b}$ will coincide. If the results are either $01$ or $10$, the corresponding bits between $k_{a}$ and $k_{b}$ will not coincide. In this last case Bob applies a bit flip. At the end of this process $k_{a} = k_{b}$.}
\end{enumerate}

Analogously as in the qutrit case, this step allows Alice and Bob to use the $n$-tuples $k_{a}$ and $k_{b}$ as the key after performing an information-reconciliation process. Once again, there is one last step regarding the security of the protocol, this will be described along with its technical details in the next subsection. Figure \ref{fig:7} shows the quantum circuit of the complete qubit-Quanty-Hall protocol.

\subsection{Security of the protocol}\label{SecBb}

In this subsection we describe the technical details in which the security of the qubit-Quanty-Hall protocol is based. To avoid introducing more notation, we will use $O$ to refer to either the pair of qubits or its measurements' results.

Notice that just from the result of the measurements of qubits $O$, Bob can know for sure to which state the system collapses after the measurement.

\begin{itemize}
	\item In the case where $O = 00$, the state that remains after the measurements is
	\begin{equation}
	\left| \phi_{b0} \right\rangle = \displaystyle\frac{1}{\sqrt{3}}  \left( \left| 001000 \right\rangle + \left| 010001 \right\rangle + \left| 100110 \right\rangle \right). \label{phib0}
	\end{equation}
	
	\item In the case where $O = 01$, the state that remains after the measurements is
	\begin{equation}
	\left| \phi_{b1} \right\rangle = \displaystyle\frac{1}{\sqrt{3}}  \left( \left| 000101 \right\rangle + \left| 011010 \right\rangle + \left| 100000 \right\rangle   \right). \label{phib1}
	\end{equation}
	
	\item In the case where $O = 10$, the state that remains after the measurements is
	\begin{equation}
	\left| \phi_{b2} \right\rangle = \displaystyle\frac{1}{\sqrt{3}}  \left( \left| 001010 \right\rangle + \left| 010000 \right\rangle + \left| 100101 \right\rangle   \right). \label{phib2}
	\end{equation}
\end{itemize}

Notice that $\left| \phi_{b0} \right\rangle$, $\left| \phi_{b1} \right\rangle$ and $\left| \phi_{b2} \right\rangle$ are all entangled states, as there is no single independent qubit. These particular states are not almost maximally entangled as a GHZ state, however they also strongly violate a Bell-type inequality for six qubits \cite{Bell1}.

For $n$ two-dimensional systems, every local hidden-variable (LHV) theory or classically correlated system must satisfy the following recursively defined inequality \cite{Bell1}:
\begin{equation}
\left| F_{n} \right| = \left|  \displaystyle\frac{1}{2} \left( a_{n} + a^{\prime}_{n} \right) F_{n-1} + \displaystyle\frac{1}{2} \left( a_{n} - a^{\prime}_{n} \right) F^{\prime}_{n-1} \right| \leq 2,  \label{Bellb}
\end{equation}
where $a_{n} = \pm 1$ and $a^{\prime}_{n} = \pm 1$ denote the two possible outcomes of two measurements on the $n$-th qubit, while $F^{\prime}_{j}$ represents the same expression as $F_{j}$ with all $a^{\prime}_{j}$ and $a_{j}$ interchanged.

For the specific case of $n=6$, notice that the value of $F_{6}$ in \eqref{Bellb} depends on the specific measurements performed in each of the six qubits, as well as on the state in which the system is. As it is usual when working with Bell-type inequalities, we calculate the value of $F_{6}$ taking the expectation value of a Bell operator $\mathcal{B}_{b}$ with respect to the state of the system \cite{Bell1}, i.e. $F_{6} = \left\langle \mathcal{B}_{b} \right\rangle $.

The respective Bell operators which maximally violate the inequality \eqref{Bellb} for the states $\left| \phi_{b0} \right\rangle$, $\left| \phi_{b1} \right\rangle$ and $\left| \phi_{b2} \right\rangle$, are found to be:
\begin{align}
\mathcal{B}_{b0} =& \, 8 \left| 101110 \right\rangle \left\langle 011001 \right| + 8 \left| 011001 \right\rangle \left\langle 101110 \right| \notag \\ 
&-8 \left| 100110 \right\rangle \left\langle 010001 \right| - 8 \left| 010001 \right\rangle \left\langle 100110 \right|,
\end{align}
\begin{align}
\mathcal{B}_{b1} =& \, 8 \left| 011010 \right\rangle \left\langle 000101 \right| + 8 \left| 000101 \right\rangle \left\langle 011010 \right| \notag \\ 
&-8 \left| 111010 \right\rangle \left\langle 100101 \right| - 8 \left| 100101 \right\rangle \left\langle 111010 \right|,
\end{align}
\begin{align}
\mathcal{B}_{b2} =& \, 8 \left| 110101 \right\rangle \left\langle 011010 \right| + 8 \left| 011010 \right\rangle \left\langle 110101 \right| \notag \\ 
&-8 \left| 100101 \right\rangle \left\langle 001010 \right| - 8 \left| 001010 \right\rangle \left\langle 100101 \right|.
\end{align}

The three of them yield a value of $\left| F_{6} \right| = 16/3 \approx 5.333$.

In this case, the ratio of violation for inequality \eqref{Bellb}, using the Bell operators $\mathcal{B}_{b0}$, $\mathcal{B}_{b1}$ and $\mathcal{B}_{b2}$ with the states $\left| \phi_{b0} \right\rangle$, $\left| \phi_{b1} \right\rangle$ and $\left| \phi_{b2} \right\rangle$ respectively, is $r = 5.333/2 \approx 2.666$; higher than the ratio of violation of the qutrit-Quanty-Hall protocol and hence higher than the $E91$ protocol.

We are now in a position to describe the last step of the qubit-Quanty-Hall protocol.

\begin{enumerate}
	\setcounter{enumi}{9}
	\item \textit{Bob measures the states $\left| \phi_{bj} \right\rangle$ and calculates the expectation value of the corresponding Bell operator $\mathcal{B}_{bj}$ in each case. If $\left\langle \mathcal{B}_{bj} \right\rangle \geq \chi$, where $\chi >  2$ is a previously agreed lower bound for $\left| F_{6} \right|$ between Alice and Bob, they conclude there was no interference in their communications and thus that the key is safe.}
\end{enumerate}

\subsection{Eve's attack} \label{Eveb}

In this subsection we describe how the presence of an spy (Eve) in the communications between Alice and Bob, might alter the results of the protocol proposed in subsection \ref{QHKDb}. We suppose that every measurement performed by Eve is a projective measurement.

Unlike the qutrit-Quanty-Hall protocol, in the qubit-Quanty-Hall protocol there is no publicly available information. This means that the only possible points of vulnerability are when Bob sends the pair of qubits $A$ to Alice and when she sends them back to Bob.

The first possible case is that Eve intercepts and measures Alice's pair of qubits the first time they are sent (step 3 of the protocol), but not the second one. In this case, as Alice has not yet applied her strategy, there is no useful information Eve could retrieve from her measurement, getting only two qubits in the state $\left| 00 \right\rangle$, $\left| 01 \right\rangle$ or $\left| 10 \right\rangle$ with a probability of $1/3$ each.

The second possible case is that Eve intercepts and measures Alice's pair of qubits the second time it is sent (step 5 of the protocol), but not the first one. Once again, due to the form of the state after the application of the INIT operator (see figure \ref{fig:1}), Eve would again only be getting two qubits in the state $\left| 00 \right\rangle$, $\left| 01 \right\rangle$ or $\left| 10 \right\rangle$ with a probability of $1/3$ each.

Just as in the qutrit-Quanty-Hall protocol, in order for Eve to have a chance of knowing which strategy Alice will apply, she has to, once again, perform two $IR$ attacks: the first one when Bob sends the qubits to Alice and the second one when Alice sends back the qubits to Bob.

We have shown that in this paradigm of projective measurements, the only possible option for Eve is to perform two different $IR$ attacks in possibly two qubits each time. We next describe the possible consequences of this action.

First notice that, if Eve measures the two qubits in pair $A$, either the first time they are sent, the second one or both, the system will collapse to a classical (not in a superposition) state, and the analysis to be made is exactly the same as in subsection \ref{Evet}. In each case, if Eve decided to perform this attack every time Alice and Bob implemented the protocol, then the possible states that remain after Bob's last measurement are respectively
\begin{align}
\hat{\varrho}_{b0} = \frac{1}{3} & \left( \left| 001000 \right\rangle \left\langle 001000 \right| + \right. \notag \\ & \left. \left| 010001 \right\rangle \left\langle 010001 \right| + \left| 100110 \right\rangle \left\langle 100110 \right| \right),
\end{align}
\begin{align}
\hat{\varrho}_{b1} = \frac{1}{3} & \left( \left| 000101 \right\rangle \left\langle 000101 \right| + \right. \notag \\ & \left. \left| 011010 \right\rangle \left\langle 011010 \right| + \left| 100000 \right\rangle \left\langle 100000 \right| \right),
\end{align}
\begin{align}
\hat{\varrho}_{b2} = \frac{1}{3} & \left( \left| 001010 \right\rangle \left\langle 001010 \right| + \right. \notag \\ & \left. \left| 010000 \right\rangle \left\langle 010000 \right| + \left| 100101 \right\rangle \left\langle 100101 \right| \right),
\end{align}
which can be shown to yield a value of $\left| F_{6} \right| = \textrm{Tr}\left\lbrace \hat{\varrho}_{bj} \mathcal{B}_{bj} \right\rbrace = 0 $ for all $j\in \left\lbrace 0,1,2 \right\rbrace $. This result, based on what was exposed in the previous subsection \ref{SecBb}, allows Alice and Bob to infer that Eve has interfered in their communications.

\begin{figure}[!t]
	\begin{centering}
		\includegraphics[width=0.9\linewidth]{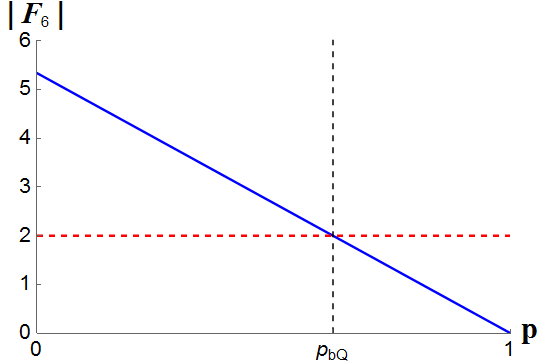}
		\par\end{centering}
	\caption{Dependence of the absolute value of the Bell parameter $F_{6}$ with the probability $p$ of Eve performing two $IR$ attacks. \label{figF6}}
\end{figure}

If now Eve decides to perform this attack based on the occurrence of an event with probability $p$, i.e. Eve's attack will be executed with probability $p$ each time the protocol is implemented, then the possible states that remain in each case after Bob's last measurement are
\begin{align}
\hat{\rho}_{0} =& \left( 1-p \right) \left| \phi_{b0} \right\rangle \left\langle \phi_{b0} \right| + p \, \hat{\varrho}_{b0}, \label{rho0b}
\\
\hat{\rho}_{1} =& \left( 1-p \right) \left| \phi_{b1} \right\rangle \left\langle \phi_{b1} \right| + p \, \hat{\varrho}_{b1}, \label{rho1b}
\\
\hat{\rho}_{2} =& \left( 1-p \right) \left| \phi_{b2} \right\rangle \left\langle \phi_{b2} \right| + p \, \hat{\varrho}_{b2}. \label{rho2b}
\end{align}
In this case $\left| F_{6} \right|$ is a function of $p$, the dependence is the same for all $\hat{\rho}_{bj}$ and its plot is shown in Figure \ref{figF6}.

Defining $p_{bQ}$ as the value of $p$ in which $\left| F_{6} \right| = 2$, from Figure \ref{figF6} we found that if $p<p_{bQ}$, the inequality \eqref{Bellb} can still be violated. The value of $p_{bQ}$ is $p_{bQ} = 0.625$, meaning that Eve can have $62.5\%$ of the raw key if Alice and Bob decided that their only criterion to consider the key as safe was the inequality violation. However, as stated in step 10 of the qubit-Quanty-Hall protocol, Alice and Bob could have previously agreed on a lower bound for $\left| F_{6} \right|$, reducing Eve's information on the key as much as they want.

A seemingly more interesting strategy for Eve to carry out is measuring just one qubit of pair $A$. In this case, if Eve measures the first qubit ($A0$) of the pair, both the first and the second time the qubits are sent, we fall again in the same case we have already analyzed, as the state collapses to a classical state. The same happens if Eve decides to measure the second qubit ($A1$) the two times the pair is sent. The only cases in which the state is not collapsed to a classical state are

\begin{itemize}
	\item Alice applied $\hat{G}_{b0}$, Bob applied $\hat{G}_{b0}$, Eve first measured $A0$ and obtained $A0 = 0$, Eve then measured $A1$ and obtained $A1 = 0$. In this case the final state of the system is
	\begin{equation}
	\left| \lambda_{00} \right\rangle = \displaystyle\frac{1}{\sqrt{2}} \left( \left| 001000 \right\rangle + \left| 010001 \right\rangle \right).
	\end{equation}
	\item Alice applied $\hat{G}_{b1}$, Bob applied $\hat{G}_{b0}$, Eve first measured $A1$ and obtained $A1 = 0$, Eve then measured $A0$ and obtained $A0 = 0$. Accordingly, the final state is given by
	\begin{equation}
	\left| \lambda_{01} \right\rangle = \displaystyle\frac{1}{\sqrt{2}} \left( \left| 011010 \right\rangle + \left| 100000 \right\rangle \right).
	\end{equation}
	\item Alice applied $\hat{G}_{b0}$, Bob applied $\hat{G}_{b1}$, Eve first measured $A0$ and obtained $A0 = 0$, Eve then measured $A1$ and obtained $A1 = 0$. Here, the final state is
	\begin{equation}
	\left| \lambda_{10} \right\rangle = \displaystyle\frac{1}{\sqrt{2}} \left( \left| 010000 \right\rangle + \left| 100101 \right\rangle \right).
	\end{equation}
	\item Alice applied $\hat{G}_{b1}$, Bob applied $\hat{G}_{b1}$, Eve first measured $A1$ and obtained $A1 = 0$, Eve then measured $A0$ and obtained $A0 = 0$. In this last case the final state is
	\begin{equation}
	\left| \lambda_{11} \right\rangle = \displaystyle\frac{1}{\sqrt{2}} \left( \left| 001000 \right\rangle + \left| 100110 \right\rangle \right).
	\end{equation}
\end{itemize}

However, even though the states $\left| \lambda_{ij} \right\rangle$ are not fully separable, they all yield a value of $\left| F_{6} \right| = 0$ for the corresponding Bell operators, and thus this case falls again in the already analyzed one.

\section{Discussion and conclusions} \label{ConclusionSection}

This work shows a possible application of quantum game theory to the area of quantum information, in particular to quantum cryptography. We have proposed two QKD protocols based on the quantum version of the Monty Hall game devised by Flitney and Abbott \cite{QMH2}. The first proposed protocol, which we referred to as the qutrit-Quanty-Hall protocol, is more directly motivated by the game, using qutrits to describe the state of the system and the same operators as in \cite{QMH2}. The motivation behind the second proposed protocol, the qubit-Quanty-Hall, was to simplify a possible physical implementation by adapting the formalism of the qutrit protocol to use qubits and simple logical quantum gates. However, in doing this adaptation, a slightly different protocol emerge. The main differences  between the two are that the qubit-Quanty-Hall protocol does not require publicly available information to be communicated between the parties, while the qutrit one does; but it does need the exchange of two particles instead of just the one needed by the qutrit protocol.

The security of both Quanty-Hall protocols relies on the non-classical correlations that arise between quantum entangled systems, just as the E91 protocol. We tested the strength of these correlations via the violation of Bell-type inequalities for two qutrits in the case of the qutrit protocol, and for six qubits in the case of the qubit protocol. The results showed that both the qutrit and the qubit protocol have a higher ratio of violation than E91, meaning that the proposed protocols have a wider margin in which the distributed key can be considered safe to use.

We have also analyzed a possible random intercept-and-resend ($IR$) attack by Eve. The results showed that, if Alice and Bob have previously agreed on a lower bound for the violation of the corresponding inequality, then they are able to arbitrarily reduce Eve's information on the key. Furthermore, we also showed that the presence of noise in the quantum channel plays against Eve, as noise by itself reduces the ratio of violation of the inequalities.

In conclusion, both proposed protocols are better entanglement-based options than the E91 protocol. The qubit-Quanty-Hall protocol in particular, is highly implementable, needing in principle just eight qubits and two quantum channels to work, and without the need of public information to be communicated.

\FloatBarrier

\acknowledgments

L. F. Quezada acknowledge support from SEP-CONACYT under project no. 288856.



\end{document}